\documentclass[twocolumn, switch]{article} 

\usepackage{preprint}

\usepackage{amsmath, amsthm, amssymb, amsfonts}
\usepackage{graphicx}
\usepackage[numbers,square]{natbib}
\usepackage[utf8]{inputenc}	
\usepackage[T1]{fontenc}	
\usepackage{xcolor}		
\usepackage[colorlinks = true,
            linkcolor = purple,
            urlcolor  = blue,
            citecolor = cyan,
            anchorcolor = black]{hyperref}	
\usepackage{booktabs} 		
\usepackage{nicefrac}		
\usepackage{microtype}		
\usepackage{lineno}		
\usepackage{float}			

\usepackage{lipsum}		

\usepackage{newfloat}
\DeclareFloatingEnvironment[name={Supplementary Figure}]{suppfigure}
\usepackage{sidecap}
\sidecaptionvpos{figure}{c}

\usepackage{titlesec}
\usepackage{tikz}
\AddEverypageHook{
	\begin{tikzpicture}[remember picture,overlay]
		\node [rotate=90, scale=1, text opacity=0.5] at ([xshift=0.5cm]current page.west) {Publication DOI: 10.1364/OL.519472};
	\end{tikzpicture}
}
\titlespacing\section{0pt}{12pt plus 3pt minus 3pt}{1pt plus 1pt minus 1pt}
\titlespacing\subsection{0pt}{10pt plus 3pt minus 3pt}{1pt plus 1pt minus 1pt}
\titlespacing\subsubsection{0pt}{8pt plus 3pt minus 3pt}{1pt plus 1pt minus 1pt}

\title{Selective excitation of high-order modes in two-dimensional cavity resonator integrated grating filters}

\usepackage{authblk}

\author[1]{Antoine Rouxel}
\author[1]{Antoine Monmayrant}
\author[1]{Stéphane Calvez}
\author[1]{Olivier Gauthier-Lafaye}

\affil[1]{LAAS-CNRS, Université de Toulouse, 31400, Toulouse}

\begin{document}

\twocolumn[ 
  \begin{@twocolumnfalse} 

\maketitle

\begin{abstract}
	The selective spatial mode excitation of a bi-dimensional grating-coupled micro-cavity called a Cavity Resonator Integrated Grating Filter (CRIGF) is reported using an incident beam shaped to reproduce the theoretical emission profiles of the device in one- and subsequently two-dimensions.
	In both cases, the selective excitation of modes up to order 10 (per direction) is confirmed by responses exhibiting one (respectively two) spectrally narrow-band resonance(s) with a good extinction of the other modes, the latter being shown to depend on the parity and order(s) of the involved modes.
	These results paves the way towards the demonstration of multi-wavelength spatially-selective reflectors or fibre-to-waveguide couplers.
	Also, subject to an appropriate choice of the materials constituting the CRIGF, this work can be extended to obtain mode-selectable laser emission or nonlinear frequency conversion.   
	
\end{abstract}
\vspace{0.35cm}

  \end{@twocolumnfalse} 
] 



\section{Cavity-resonator-integrated grating filters}

\begin{figure}[ht]
	\centering
	\includegraphics[width=0.8\linewidth]{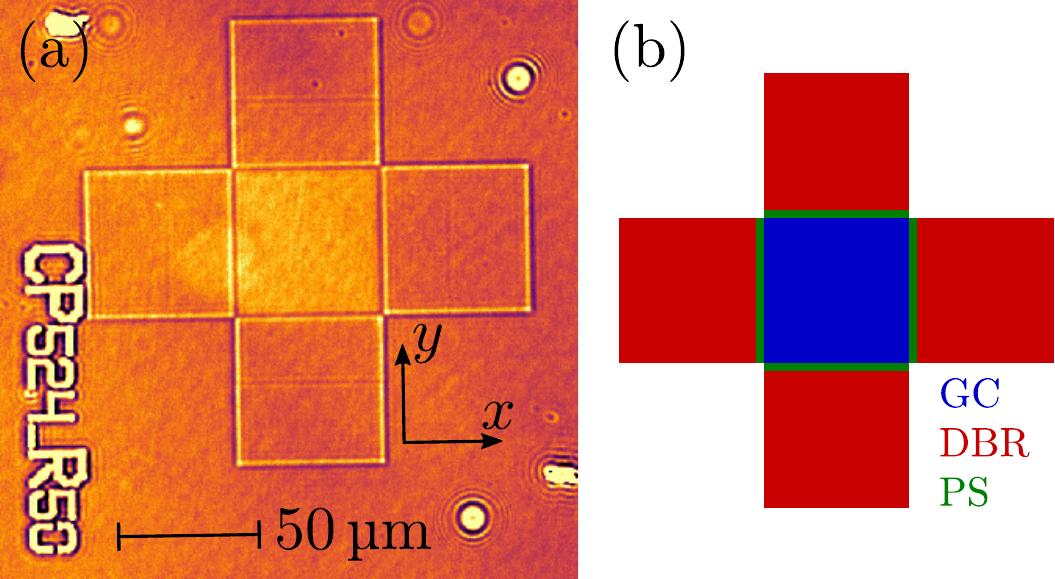}
	\caption{
		(a) Top view of the 2D CRIGF under study;
		(b) Schematic of the 2D CRIGF with a 2D central grating coupler (GC),
		surrounded by four phase sections (PS) and
		four distributed Bragg Reflectors (DBR).
		Incident light coupled by the GC is trapped in the 2D Fabry-Pérot cavity formed by the DBRs.
	}
	\label{fig:CRIGF_structure}
\end{figure}

Cavity-Resonator-Integrated Grating Filters (CRIGFs) introduced in 2012\,\cite{Kintaka2012} are a particular variant of Guided Mode Resonance Filters (GMRFs), acting as spectrally selective filters for tightly focused beams under normal incidence.
They rely on a single-mode waveguide capped with several gratings.
Usual CRIGFs are made of 1D gratings, and exhibit a strong polarization dependence.
2D-CRIGFs can provide polarization-independent reflectivity \cite{Kintaka2012b}.
As depicted in figure~\ref{fig:CRIGF_structure}, a 2D CRIGF is formed by a vertical ($y$) and an horizontal ($x$) CRIGF sharing a common central area (blue).
Along both $x$ and $y$ directions, a pair of Distributed Bragg Reflector (DBR) forms a planar Fabry-Pérot cavity (FP) for the guided mode.
A few-period-long Grating Coupler (GC), centred in the FP couples the cavity mode to focused incident beam(s).
Along each direction, a Phase Section (PS) ensures spatial and spectral overlap of the GC with the cavity mode.
Polarization independence is ensured when both the vertical and horizontal CRIGFs forming the 2D CRIGF have the same geometry: the horizontal and vertical polarization components of the incident beam are reflected respectively by the vertical and horizontal CRIGFs that exhibit identical spectral reflectivities.

In this work, we use such CRIGFs, as depicted on fig.~\ref{fig:CRIGF_structure}, made of a 2D square lattice GC with $N_{GC}=101$ periods of $\Lambda_{GC}=524\,$nm, and four DBRs with $N_{DBR}=200$ periods of $\Lambda_{DBR}=262\,$nm.
The central GC thus forms a $L \times L$ square with $L = N_{GC}\Lambda_{GC} \simeq 53\,$µm.
A more detailed description of the sample can be found in \cite{Laberdesque2015}.

\begin{figure}[ht]
	\centering
	\includegraphics[width=1\linewidth]{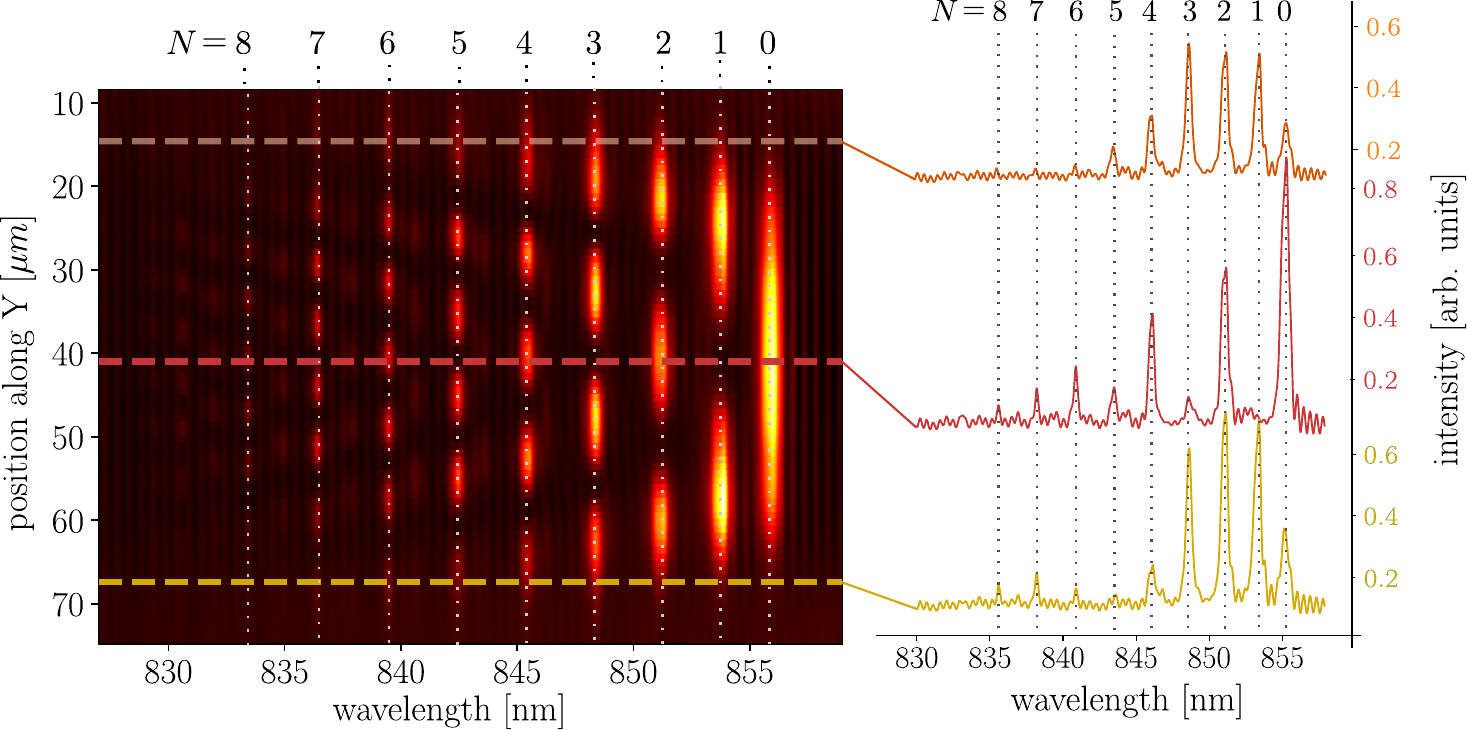}
	\caption{Spatio-spectral map of reflectivity exhibiting the higher-order radiative patterns of the cavity along the $y$ direction. The vertical dashed lines correspond to the wavelengths of the corresponding modes.
	}
	\label{fig:gaussian_cavity_scanning}
\end{figure}

Previous work~\cite{Laberdesque2015} has demonstrated the existence of multiple-order radiative patterns in 1D \& 2D CRIGFs.
As evidenced by modelling,
these radiative patterns results from the interaction of the GC with the various Fabry-Pérot modes supported by the high-reflectivity stopband of the DBRs. 
Using $x$-polarized incident light, these radiative patterns related to the vertical cavity (along the $y$ direction) of the 2D CRIGF are revealed by performing a spatial scan of the structure along the direction $y$ with a Gaussian spot just a few micrometers in diameter. 
At each position, the spectral reflectivity is measured.
The resulting spectro-spatial reflectivity map presented on figure~\ref{fig:gaussian_cavity_scanning} exhibits
several reflectivity peaks whose intensity varies with the $y$ position.
The spatial profile associated with each spectral peak presents a number $N$ of nodes along the $y$ direction that increases as the excitation wavelength decreases.
As shown in reference ~\cite{Laberdesque2015}, these profiles directly correspond to the intensity profile of the radiated field for each eigenmode of the CRIGF.

\section{Mode-selective excitation in 1D}

Experimental mode-selective excitation is accomplished using a Fourier-transform optical setup, incorporating a phase-only Spatial Light Modulator (SLM) ~\cite{Golan2009}.
As detailed in~\cite{Davis1999}, by spatially modulating the amplitude of a blazed grating displayed on the SLM, this setup enables amplitude and phase information to be encoded onto the laser beam.
More specifically, we use this setup to encode beams with profiles that correspond to the radiative field spatial distribution radiated by the CRIGF.
We start with 1D profiles to only excite the vertical cavity of the 2D CRIGF.
Each vertical pattern $M_l$, where $l$ denotes the number of nodes along the $y$ direction, has a spatial amplitude profile $A_l(y)$ given by :
\begin{equation}
	\begin{gathered}
		A_{l}(x,y) = \text{rect}\!\left(\frac{x}{L}, \frac{y}{L}\right)\times
		\begin{cases}  
			\cos\left(\pi y(l+1)/L\right) & \text{if } l \text{ is even} \\
			\sin\left(\pi y(l+1)/L\right) & \text{otherwise}
		\end{cases} \\
	\end{gathered}
	\label{eq:profile}
\end{equation}

\begin{figure}[ht]
	\centering
	\includegraphics[width=1\linewidth]{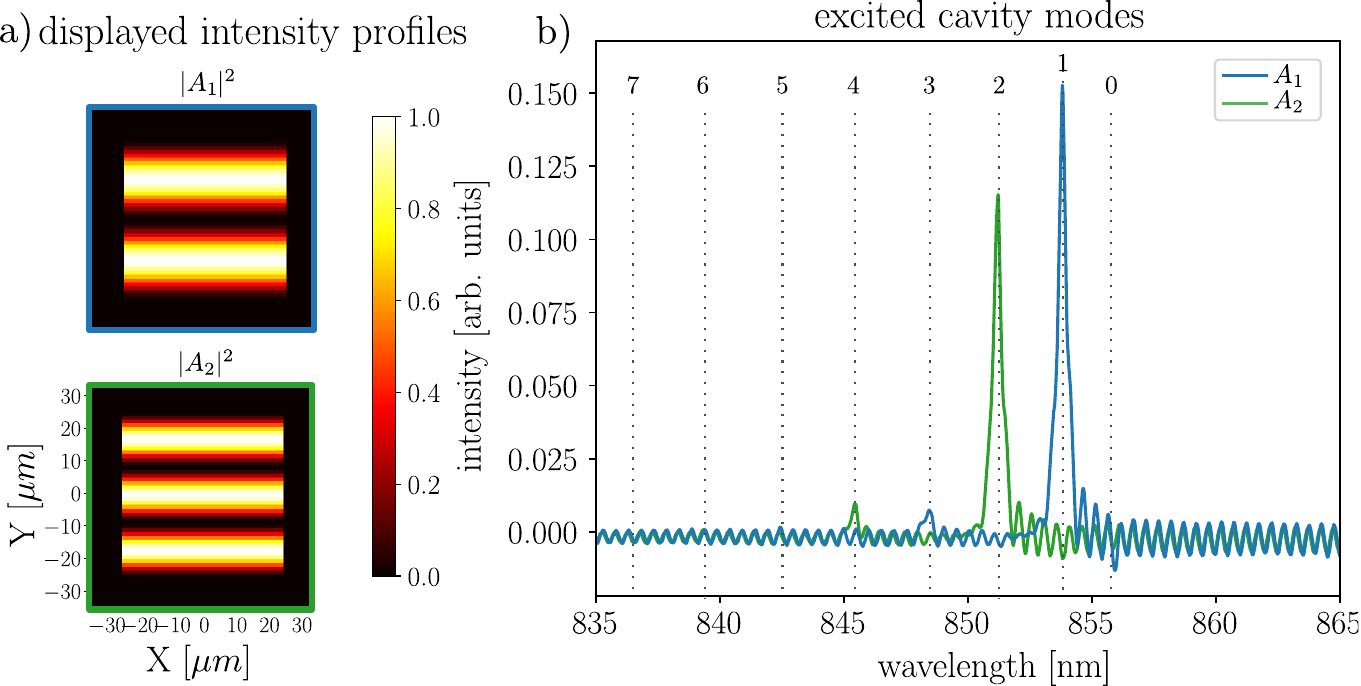}
	\caption{Selective excitation of the high-order cavity modes along the $y$ direction.
	}
	\label{fig:selective_excitation_1D}
\end{figure}

On figure~\ref{fig:selective_excitation_1D}-a), we show two examples of shaped incident beam profiles corresponding respectively to the higher-order modes $M_1$ and $M_2$ of the cavity.
In practice, the laser beam was polarized along the $x$ direction, centred on the cavity and shaped along the $y$ direction according to Eq.\,\ref{eq:profile}.
Experimental spectra for these profiles, presented in figure~\ref{fig:selective_excitation_1D}-b), are the reflectivity measurements subtracted by the thin-film response on the DBR section. 
As can be observed on these spectra, each mode reflectivity spectrum is essentially reduced to a single strong reflectivity peak, whose wavelength corresponds to the resonant wavelength of the CRIGF mode with the same spatial amplitude distribution (fig.\,\ref{fig:gaussian_cavity_scanning}).
On each spectrum,
one can also see oscillations around the baseline that are due to parasitic reflections on the back-side of the substrate
and a small peak at mode order $N+2$, which we will discuss subsequently.
The emergence of strong intensity peaks at wavelengths matching the targeted modes confirms the efficacy of the selective excitation.

\begin{figure}[ht]
	\centering
	\includegraphics[width=0.85\linewidth]{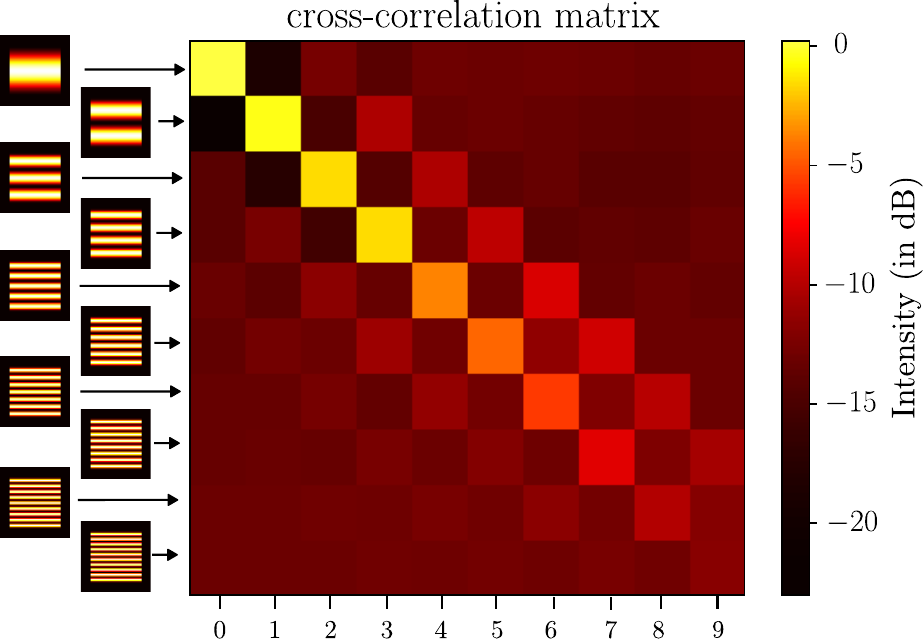}
	\caption{Cross-correlation matrix representing the excitation/attenuation of the cavity modes (columns) for the corresponding spatial profiles.}
	\label{fig:cross_corrrelation_matrix}
\end{figure}

We repeated these measurements for modes $M_0$ up to mode $M_9$. 
From these measurements we extracted the cross-correlation matrix depicted in figure~\ref{fig:cross_corrrelation_matrix}, in which each square colour depicts the reflectivity at a given resonant mode wavelength (columns) for a given excitation spatial profile (lines).
For a clearer picture, we choose to normalize the matrix to the fundamental mode response (reference in the matrix top-left corner), and display the responses in logarithmic scale, hereby presenting the modal excitation selectivity.
As anticipated, the highest values are along the matrix diagonal, indicating predominant excitation of the targetted modes.
Notably, the intensity of the reflected signal diminishes for higher modes.
Additionally, a chequerboard pattern is observed in this cross-correlation matrix, explainable by mode parity: even-numbered modes, when targeted, also partially excite other even-numbered modes due to spatial profile overlap, as shown in figure\,\ref{fig:cross_corrrelation_matrix}. 
A similar pattern occurs for odd-numbered modes.


The experimental results demonstrate that selective excitation of 1D modes is achievable up to mode $M_9$ for the $y$ direction, offering promising avenues for experimental modal decomposition of beams.
Note that for all these measurements, we applied the theoretical profiles defined in \eqref{eq:profile}, without any optimisation of either the central position or the periodicity.

\section{Mode selective excitation in 2D}

As a further proof of concept, we also recorded the reflectivity from the same 2D-CRIGF when excited by a beam polarized at $45$\textdegree\ and shaped along both $x$ and $y$ directions.
In that case, both the vertical and horizontal modes are simultaneously excited with an efficiency that respectively depends on the spatial profiles along $x$ and $y$ directions.
We refer to these modes as $M_{k,l}$ with 2D profiles $A_{k,l}(x,y)$ defined as:
\begin{equation}
	A_{k,l}(x,y) = A_{k}(x,y)\times A_{l}(y,x)
	\label{eq:2Dprofile}
\end{equation}

\begin{figure}[ht]
	\centering
	\includegraphics[width=1\linewidth]{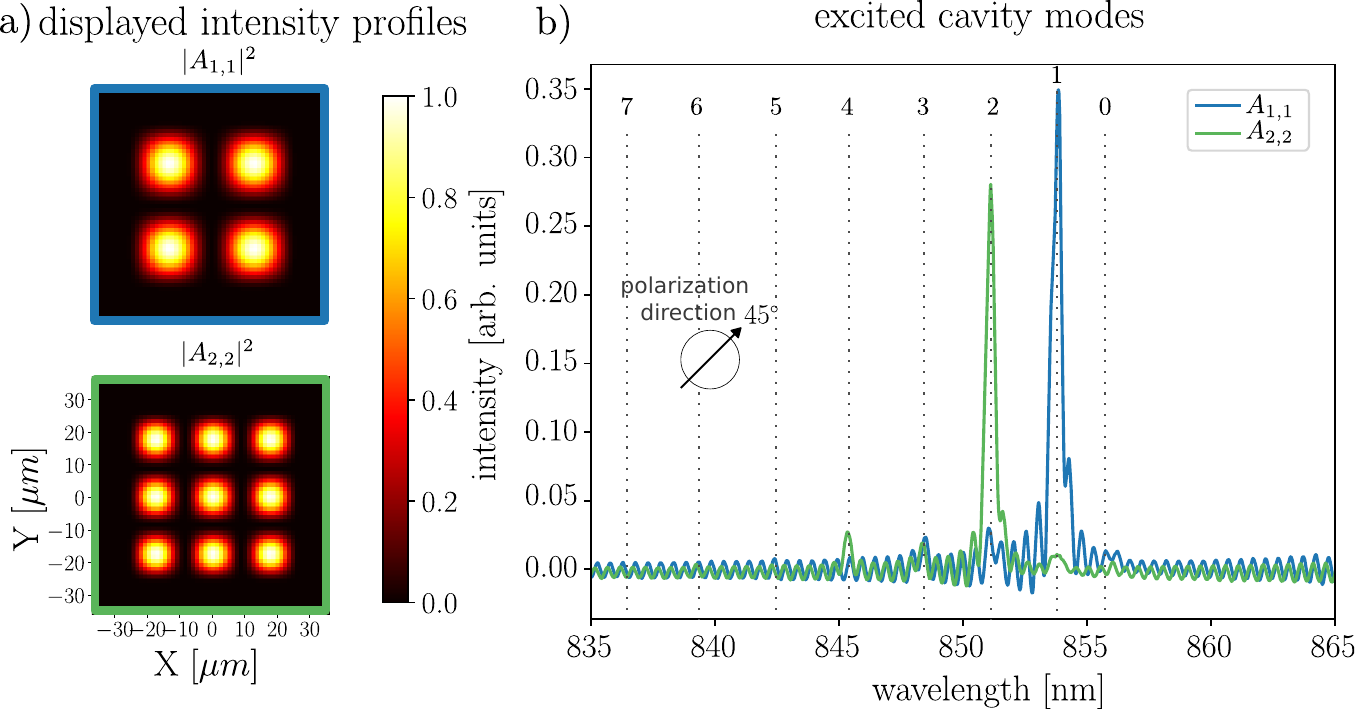}
	\caption{Selective excitation of the 2D cavity with symmetric 2D shaped beams, for 45° polarization of the excitation beam.
	}
	\label{fig:sym_2Dmodes}
\end{figure}

Figure\,\ref{fig:sym_2Dmodes} shows the reflectivity spectra obtained under illumination by an $A_{1,1}$ and an $A_{2,2}$ beam profiles.
Each profile exhibits the same number of nodes along both $x$ and $y$ directions.
These spectra are extremely similar to that presented in figure\,\ref{fig:selective_excitation_1D} except that both vertical and horizontal cavities of the 2D CRIGF are now simultaneously excited.
Indeed, as the vertical and horizontal cavities in the 2D CRIGF are identical, for each excitation profile with an identical number of nodes along the  $x$ and $y$ directions, the spectrum exhibits one single peak corresponding to the wavelength-degenerated signatures of both the vertical and the horizontal cavities, both cavities emitting along orthogonal polarizations.

\begin{figure}[ht]
	\centering
	\includegraphics[width=1\linewidth]{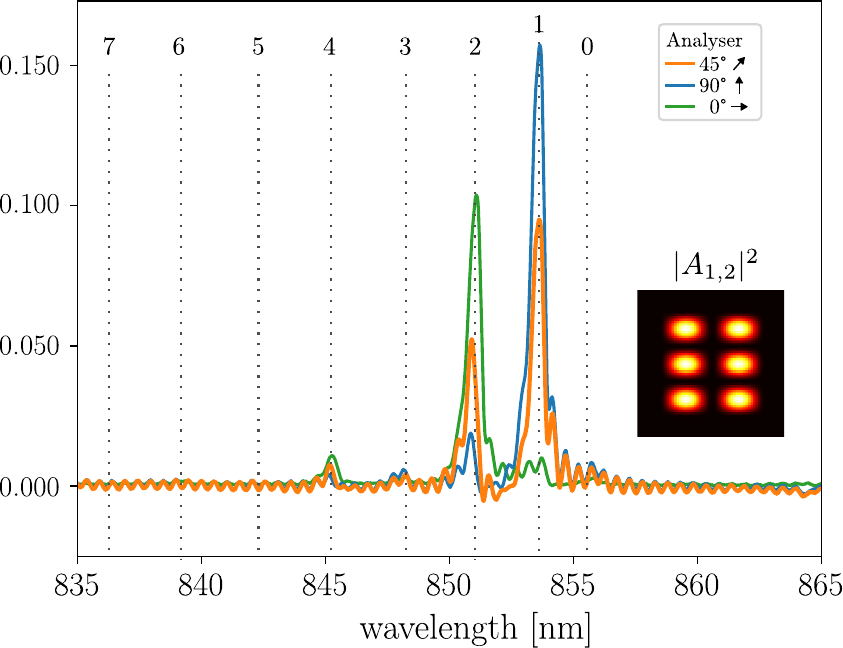}
	\caption{Selective excitation of the 2D cavity with asymmetric 2D shaped beams, polarized at 45°, for different orientation of the analyser placed before the photodiode collecting the reflected light.
	}
	\label{fig:asym_2Dmode}
\end{figure}

Conversely, figure\,\ref{fig:asym_2Dmode} shows the polarized reflectivity spectra obtained under a 45°-polarized illumination with a different number of nodes along the $x$ and $y$ directions, here chosen to be an $A_{1,2}$ beam profile.
For a 45°-polarized analyser placed in front of the photodiode monitoring the sample reflection, we observe the whole signal emitted by the 2D CRIGF (orange line).
As the excited modes along the vertical and the horizontal directions are different,  two distinct peaks can be observed in the spectral signature, one for each cavity.
This is further confirmed by analysing the reflectivity response for other orientations of the analyser. For an horizontally polarized analyser, only the contribution from the vertical CRIGF can be detected. Here, only the shorter wavelength peak is observed (green line), in agreement with the fact that the highest-order of the two modes is excited in the vertical CRIGF (with the $A_2(y)$ profile).
Similarly, for a vertically-oriented analyser, we only observe the contribution from the horizontal cavity and it corresponds to the longer wavelength peak (blue line), excited by the $A_1(x)$ profile along the horizontal direction.

The latter demonstration proves that we are able to selectively excite two different modes in (2D) crossed CRIGFs, thereby paving the way towards 2D beam spatial decomposition and showing an ability to treat (combine) a much higher number of modes (having shown here the capability to control of up to $10\!\times\!10$-modes) while maintaining a more compact format than would be achievable with the 1D version.

\section{Conclusion}
In this manuscript, we have demonstrated both one- and two-dimensional selective excitation of the high-order modes supported by a Cavity-Resonator-Integrated Grating Filter (CRIGF), using shaped excitation beams.

This selective excitation is obtained by shaping the incident beam so as to reproduce the theoretical emission profiles of the CRIGF.
For 1D-shaped beams, each incident beam profile selectively excites a spectrally narrow-band resonance along the (polarization-)chosen direction of the structure while, in the 2D case, each incident beam profile excites two narrow-band resonances, one along each direction in the plane of the structure, these two resonances being degenerated or not.
The achieved selectivity was shown to depend on the parity of modes involved and to decrease as the order of the mode(s) increases.
Some improvement could be achieved by further optimizing the incident beam profiles, to compensate for slight experimental deviation from theory (profile periodicity, profile centring, etc...).

Future work could involve the use of these mode-discriminating spectral reflectors as end mirrors in a multimode fibre laser as an alternative to the combination of few-mode fibre Bragg grating and mode-selective couplers used in \cite{Yao2018}.

Furthermore, this dual-mode excitation at different wavelengths could be performed on nonlinear CRIGFs \cite{Renaud2019} with a view to explore frequency-mixing processes with various phase-matching geometries and wavelength separation for laser wavelength spectral shifting or the generation of entangled photon pairs by spontaneous parametric down conversion \cite{Maltese2020}.

Alternatively, since CRIGFs are derived from structures that were originally designed for efficient waveguide in-/output coupling \cite{Kintaka2010,OL10-35-12-1989-1991}, one could design shape- and wavelength-selective directional couplers to bridge the gap between Photonic Integrated Circuits (PICs) and few mode fibres.
Contrarily to recently proposed couplers for few mode fibres \cite{Zhang2018,Cheng2021}, these CRIGF-based couplers would be both shape- and wavelength-selective and a single structure could selectively handle many more  than two different modes.

\bibliography{main}


\end{document}